\documentclass[twocolumn,twosides,letterpaper,10pt]{IEEEtran}
\usepackage{cite,url,color}
\usepackage{graphicx} 
\graphicspath{{./pdf/}{./jpeg/}{./figs/}{./eps/}}
\DeclareGraphicsExtensions{.pdf,.jpeg,.png,.eps}
\usepackage[cmex10]{amsmath}
\usepackage{amssymb,bm}
\usepackage{algpseudocode} 
\newcommand{\algrule}[1][.5pt]{\par\vskip.25\baselineskip\hrule height #1\par\vskip.25\baselineskip}
\usepackage[caption=false,font=footnotesize]{subfig}
\usepackage[amsmath,thmmarks,hyperref]{ntheorem}

\usepackage{multirow}
\newcommand{\ve}[1]{\mathbf{{#1}}}
\newcommand{\abs}[1]{\ensuremath{\vert #1\vert}}
\newcommand{\norm}[1]{\ensuremath{\Vert #1\Vert}}

\begin{document}
%
\title{Fast Marginalized Block Sparse Bayesian Learning Algorithm}
%
%
%

\author{Benyuan~Liu$^{1}$,~Zhilin~Zhang$^{2}$,~Hongqi~Fan$^1$,~Qiang~Fu$^1$%
\thanks{$^1$Benyuan Liu, Hongqi Fan and Qiang Fu are with The Science and Technology on Automatic Target Recognition Laboratory, National University of Defense Technology. Changsha, Hunan, P. R. China, 410074. E-mail: liubenyuan@gmail.com}%
\thanks{$^2$Zhilin Zhang is with The Emerging Technology Lab, Samsung R\&D Institute America - Dallas, 1301 E. Lookout Drive, Richardson, TX 75082, USA. E-mail: zhilinzhang@ieee.org}%
\thanks{B. Liu and H. Fan were supported in part by the National Natural Science Foundation of China under Grant 61101186.}%
\thanks{This is a technical report. An extended version was submitted to IEEE Journal of Biomedical and Health Informatics with the title ``Energy Efficient Telemonitoring of Physiological Signals via Compressed Sensing: A Fast Algorithm and Systematic Evaluation''.}}

%
%

\markboth{Technical Report.}%
{} 
%



\maketitle
\begin{abstract}
The performance of sparse signal recovery from noise corrupted, underdetermined measurements can be improved if both sparsity and correlation structure of signals are exploited. One typical correlation structure is the intra-block correlation in block sparse signals. To exploit this structure, a framework, called block sparse Bayesian learning (BSBL), has been proposed recently. Algorithms derived from this framework showed superior performance but they are not very fast, which limits their applications. This work derives an efficient algorithm from this framework, using a marginalized likelihood maximization method. Compared to existing BSBL algorithms, it has close recovery performance but is much faster. Therefore, it is more suitable for large scale datasets and applications requiring real-time implementation.
\end{abstract}


\ifCLASSOPTIONpeerreview
\fi
%
\IEEEpeerreviewmaketitle

%
%
\section{Introduction}
Sparse signal recovery and the associated compressed sensing \cite{Candes2008a} can recover a signal with small number of measurements with high probability of successes (or sufficient small errors), given that the signal is sparse or can be sparsely represented in some domain. It has been found that exploiting structure information\cite{Baraniuk2010,Zhang2012a} of a signal can further improve the recovery performance. In practice, a signal generally has rich structures. One structure widely used is the block/group sparse structure\cite{Zhang2012a,Yuan2006,Baraniuk2010,Babacan2012}, which refers to the case when nonzero entries of a signal cluster around some locations. Existing algorithms exploit such information showed improved recovery performance.

Recently, noticing intra-block correlation widely exists in real-world signals, Zhang and Rao \cite{Zhang2012a,Zhang2011} proposed the block sparse Bayesian learning (BSBL) framework. A number of algorithms have been derived from this framework, and showed superior ability to recover block sparse signals or even non-sparse signals \cite{Zhang_TBME2012b}. But these BSBL algorithms are not fast, and thus cannot be applied to large-scale datasets.

In this work, we propose an efficient implementation using the fast marginalized likelihood maximization (FMLM) method \cite{Tipping2003}. Thanks to the BSBL framework, it can exploit both block structure and intra-block correlation. Experiments conducted on both synthetic data and real life data showed that the proposed algorithm significantly outperforms traditional algorithms which only exploit block structure such as Model-CoSaMP\cite{Baraniuk2010} and Block-OMP\cite{Eldar2010}. It has similar recovery accuracy as BSBL algorithms \cite{Zhang2012a}. However, it is much faster than existing BSBL algorithms\cite{Zhang2012a} and thus is more suitable for large scale problems. 

Throughout the paper, {\bf Bold} symbols are reserved for vectors $\ve{a}$ and matrices $\ve{A}$. $\mathrm{Tr}(\ve{A})$ computes the trace of the matrix. $\mathrm{diag}(\ve{A})$ extracts the diagonal vector from a matrix $\ve{A}$ and $\mathrm{diag}^{-1}(\ve{a})$ builds a matrix with $\ve{a}$ as its diagonal vector. $\ve{A}^T$ denotes the transpose of matrix $\ve{A}$.

%
%
\section{The Framework of the Block Sparse Bayesian Learning Algorithm}
%

\subsection{The basic BSBL Framework\cite{Zhang2012a}}
A block sparse signal $\ve{x}$ has the following structure:
\begin{equation}
\ve{x} = [\underbrace{x_1,\cdots,x_{d_1}}_{\ve{x}_1^T},\cdots,
\underbrace{x_1,\cdots,x_{d_g}}_{\ve{x}_g^T}]^T,
\end{equation}
which means $\ve{x}$ has $g$ blocks, and only a few blocks are nonzero. Here $d_i$ is the block size for the $i$th block. Moreover, for time series data $\ve{x}$, the samples within each block are usually correlated. To model the block sparse and intra-block correlation, the BSBL framework\cite{Zhang2012a} suggests to use the parameterized Gaussian distribution:
\begin{equation}
p(\ve{x}_i;{\gamma_i},\ve{B}_i) =
\mathcal{N}(\ve{x}_i;\mathbf{0},{\gamma_i}\ve{B}_i). \label{eq:x_prior}
\end{equation}
with unknown deterministic parameters $\gamma_i$ and $\ve{B}_i$. $\gamma_i$ represents the confidence of the relevance of the $i$th block and $\ve{B}_i$ captures the intra-block correlation. The framework further assumes that blocks are mutually independent. Therefore, we write the signal model as,
\begin{equation}
    p(\ve{x};\{\gamma_i\},\{\ve{B}_i\}) = \mathcal{N}(\ve{x};\ve{0},\bm{\Gamma}),
\end{equation}
where $\bm{\Gamma}$ is a block diagonal matrix with the $i$th principal diagonal given by $\gamma_i\ve{B}_i$. 

The observation $\ve{y}$ is obtained by
\begin{equation}
\ve{y} = \bm{\Phi}\ve{x} + \ve{n},
\end{equation}
where $\bm{\Phi}$ is a $M\times N$ sensing matrix and $\ve{n}$ is the observation noise. The sensing matrix $\bm{\Phi}$ is an underdetermined matrix and the observation noise is assumed to be independent and Gaussian with zero mean and
variance equal to $\beta^{-1}$. $\beta$ is also unknown. Thus the likelihood is given by
\begin{equation}
p(\ve{y}|\ve{x};\beta) = \mathcal{N}(\bm{\Phi}\ve{x},\beta^{-1}\ve{I}). \label{eq:y_prior}
\end{equation}

The main body of the BSBL algorithm iteratively between the estimation of the posterior 
$p(\ve{x}|\ve{y}; \{\gamma_i,\mathbf{B}_i\},\beta)=\mathcal{N}(\bm{\mu},\bm{\Sigma})$
with
$\bm{\Sigma} \triangleq (\bm{\Gamma}^{-1} + \bm{\Phi}^T\beta\bm{\Phi})^{-1}$ and
$\bm{\mu} \triangleq \bm{\Sigma}\bm{\Phi}^T\beta\ve{y}$ and 
maximization the likelihood $p(\ve{y}|\{\gamma_i,\ve{B}_i\},\beta)=\mathcal{N}(\ve{y};\ve{0},\ve{C})$ with 
$\ve{C} = \beta^{-1}\ve{I} + \bm{\Phi}\bm{\Gamma}\bm{\Phi}^T$. The update rules for the parameters 
$\{\gamma_i,\ve{B}_i\}$ and $\beta$ are derived using the Type II Maximum Likelihood\cite{Tipping2001,Zhang2012a} method,
which leads to the following cost function,
\begin{equation}
\mathcal{L}(\{\gamma_i,\mathbf{B}_i\},\beta)
 = \log\abs{\ve{C}} + \ve{y}^T\ve{C}^{-1}\ve{y}, \label{eq:log-likelihood}
\end{equation}
Once all the parameters, namely $\{\gamma_i,\mathbf{B}_i\},\beta$,  are estimated, the MAP estimate of the signal $\mathbf{x}$ can be directly obtained from the mean of the posterior, i.e.,
\begin{equation}
\mathbf{x} = \bm{\Sigma}\bm{\Phi}^T\beta\ve{y}. \label{eq:Rule_x}
\end{equation}

\subsection{The Extension to the BSBL Framework}
In the original BSBL framework\cite{Zhang2012a}, $\gamma_i$ and $\ve{B}_i$ together formed the covariance matrix of the $i$th block $\ve{x}_i$, which can be conveniently modeled using a symmetric, positive semi-definite matrix $\ve{A}_i$,
\begin{equation}
    p(\ve{x}_i; \ve{A}_i) = \mathcal{N}(\ve{x}_i; \ve{0}, \ve{A}_i).
\end{equation}

The diagonal element of $\ve{A}_i$, denoted by $A_i^{jj}$, models the variance of the $j$th signal $x_i^j$ in the $i$th block $\ve{x}_i$. Such variance parameter $A_i^{jj}$ determines the relevance\cite{Tipping2001} for the signal $x_i^j$. In the extension to the BSBL framework, we may conveniently introduce the term \emph{block relevance}, defined as the average of the estimated variance of signals in $\ve{x}_i$,
\begin{equation}\label{eq:block_relevance}
	\gamma_i \triangleq \frac{1}{d_i}\mathrm{Tr}(\ve{A}_i).
\end{equation}
The notation \emph{block relevance} cast the Relevance Vector Machine\cite{Tipping2001} as a specialized form of one BSBL variant where we assume $\ve{A}_i$ has the following structure,
\begin{equation}
	\ve{A}_i = \mathrm{diag}^{-1}(A_i^1,\cdots,A_i^{d_i}).
\end{equation}
This variant of BSBL is for the signals with block structure and each signal in the blocks are spiky and do not correlate with each other. 

The off-diagonal elements $\ve{A}_i^{jk}, j\neq k$ models the covariance of the block signal. It has been shown that exploit such structural information is beneficial in recovering piecewise smooth block sparse signals\cite{Zhang2012a,Zhang_TBME2012b}. There are rich classes of covariance matrix, namely Compound Symmetric, Auto-Regressive (AR), Moving-Average (MA) etc. These structures can be inferred during the learning process of $\ve{A}_i$. 

The signal model and the observation model are built up in a similar way,
\begin{align}
    p(\ve{x};\{ \ve{A}_i \}) &= \mathcal{N}(\ve{x}; \ve{0}, \bm{\Gamma}), \\
    p(\ve{y};\{ \ve{A}_i \}, \beta) &= \mathcal{N}(\ve{y}; \bm{\Phi}\ve{x}, \beta^{-1}\ve{I}).
\end{align}
The parameters $\{\ve{A}_i\}$ and $\beta$ can be estimated from the cost function \eqref{eq:log-likelihood} using the type II maximization method.

\section{The Fast Marginalized Block SBL Algorithm}
There are several methods to minimize the cost function \eqref{eq:log-likelihood}. In \cite{Zhang2012a}, the author provided a bound optimization method and a hybrid $\ell_1$ method to derive the update rules for $\gamma_i$ and $\ve{B}_i$. 
In the following we consider to extend the marginalized likelihood maximization method within the BSBL framework. This method was used by Tipping et al. \cite{Tipping2003} for their fast SBL algorithm and later by Ji et al. \cite{Ji2008} for their Bayesian compressive sensing algorithm.

\subsection{The Main-body of the Algorithm}
The cost function \eqref{eq:log-likelihood} can be optimized in a block way. 
We denote by $\bm{\Phi}_i$ the $i$th block in $\bm{\Phi}$ with the column indexes corresponding to the $i$th block of the signal $\ve{x}$. Then $\ve{C}$ can be rewritten as:
\begin{align}
\ve{C} &= \beta^{-1}\ve{I} + \sum_{m\neq i} \bm{\Phi}_m\ve{A}_m\bm{\Phi}_m^T+
\bm{\Phi}_i\ve{A}_i\bm{\Phi}_i^T, \\
 &= \ve{C}_{-i} + \bm{\Phi}_i\ve{A}_i\bm{\Phi}_i^T, \label{eq:c}
\end{align}
where $\ve{C}_{-i} \triangleq \beta^{-1}\ve{I} + \sum_{m\neq i} \bm{\Phi}_m\ve{A}_m\bm{\Phi}_m^T$.
Using the Woodbury Identity,
\begin{align}
\abs{\ve{C}} &= \abs{\ve{A}_i}\abs{\ve{C}_{-i}}\abs{\ve{A}_i^{-1} + \ve{s}_i}, \\
\ve{C}^{-1} &= \ve{C}_{-i}^{-1} - \ve{C}_{-i}^{-1}\bm{\Phi}_i(\ve{A}_i^{-1} + \ve{s}_i)^{-1}\bm{\Phi}_i^T\ve{C}_{-i}^{-1},
\end{align}
where
$\ve{s}_i\triangleq \bm{\Phi}_i^T\ve{C}_{-i}^{-1}\bm{\Phi}_i$ and
$\ve{q}_i\triangleq\bm{\Phi}_i^T\ve{C}_{-i}^{-1}\ve{y}$.
Equation \eqref{eq:log-likelihood} can be rewritten as:
\begin{align}
\mathcal{L} =& \log\abs{\ve{C}_{-i}} + \ve{y}^T\ve{C}_{-i}^{-1}\ve{y} \nonumber \\
 &+ \log\abs{\ve{I}_{d_i} + \ve{A}_i\ve{s}_i} - \ve{q}_i^T(\ve{A}_i^{-1} + \ve{s}_i)^{-1}\ve{q}_i, \\
 =& \mathcal{L}(-i) + \mathcal{L}(i),
\end{align}
where $\mathcal{L}(-i) \triangleq \log\abs{\ve{C}_{-i}} + \ve{y}^T\ve{C}_{-i}^{-1}\ve{y}$, and
\begin{equation}\label{eq:lgi}
\mathcal{L}(i) =
\log\abs{\ve{I}_{d_i} + \ve{A}_i\ve{s}_i} -
\ve{q}_i^T(\ve{A}_i^{-1} + \ve{s}_i)^{-1}\ve{q}_i,
\end{equation}
which only depends on $\mathbf{A}_i$.


Setting $\frac{\partial \mathcal{L}(i)}{\partial \ve{A}_i}=\ve{0}$, we have the updating rule
\begin{equation}\label{eq:a_0}
\ve{A}_i = \ve{s}_i^{-1}(\ve{q}_i\ve{q}_i^T - \ve{s}_i)\ve{s}_i^{-1}.
\end{equation}

The block relevance $\gamma_i$ is calculated using \eqref{eq:block_relevance} and the correlation structural is inferred from \eqref{eq:a_0} by investigating a symmetric matrix $\ve{B}_i$ calculated as
\begin{equation}
	\ve{B}_i = \frac{\ve{A}_i}{\gamma_i}.
\end{equation}
The diagonal elements of $\ve{B}_i$ are regularized to unity (i.e., $\mathrm{diag}(\ve{B}_i) = \ve{1}$) to maintain the spectrum of $\ve{A}_i$.

%
%
\subsection{Imposing the Structural Regularization on $\mathbf{B}_i$}

As noted in \cite{Zhang2012a},  regularization to $\mathbf{B}_i$ is required due to limited data. It has been shown \cite{Zhang2011}  that in noiseless cases the regularization does not affect the global minimum of the cost function (\ref{eq:log-likelihood}), i.e., the global minimum still corresponds to the true solution; the regularization only affects the probability of the algorithm to converge to the local minima. A good regularization can largely reduce the probability of local convergence. Although theories on regularization strategies are lacking, some empirical methods \cite{Zhang2011,Zhang2012a} were presented. 

The proposed algorithm is extensible in that it can incorporate different time-series correlation models. In this paper we focus on the following forms of correlation models.

\subsubsection{Simple (SIM) Correlation Model}
In the simple (SIM) correlation model, we fix $\ve{B}_i=\ve{I}(\forall i)$ and ignores the correlation within the signal block. Such regularization is appropriate for recovering the signal in the transformed domain, i.e., via Fourier or Discrete Cosine transform. In these cases, the coefficients are spiky sparse and may cluster into a few non-zero blocks.
Our algorithm using this regularization is denoted by {\bf BSBL-FM(0)}. 

\subsubsection{Auto-Regressive (AR) Correlation Model}
We model the entries in each block as an AR(1) process with the coefficient $r_i$. As a result,
$\ve{B}_i$ has the following form
\begin{equation}
\ve{B}_i = \mathrm{Toeplitz}([1,r_i,\cdots,r_i^{d_i-1}]). \label{equ:Bi}
\end{equation}
where $\mathrm{Toeplitz}(\cdot)$ is a MATLAB command expanding a real vector into a
symmetric Toeplitz matrix. Thus the correlation level of the intra-block correlation is reflected by the value of $r_i$. 
$r_i$ is empirically calculated\cite{Zhang2012a} by $r_i \triangleq\frac{m_1^i}{m_0^i}$,
where $m_0^i$ (res. $m_1^i$) is the average of entries along the main diagonal
(res. the main sub-diagonal) of the matrix $\ve{B}_i$. This calculation cannot ensure $r_i$ has a feasible value, i.e. $|r_i|<0.99$. Thus in practice, we calculate $r_i$ by
\begin{align}
r_i &= \mathrm{sign}(\frac{m^i_1}{m^i_0}) \min \Big\{\Big|\frac{m^i_1}{m^i_0}\Big|,0.99 \Big\}. \label{equ:r_i}
\end{align}
Our algorithm using this regularization (\ref{equ:Bi})-(\ref{equ:r_i}) is denoted by {\bf BSBL-FM(1)}.

\subsubsection{Additional Average Step}
In many real-world applications, the intra-block correlation in each block of a signal
tends to be positive and high together. Thus, one can further constrain that
all the intra-block correlation of blocks have the same AR coefficient $r$ \cite{Zhang2012a}, 
\begin{align}
r &= \frac{1}{g}\sum_{i=1}^g r_i,  \label{equ:r}
\end{align}
where $r$ is the average of $\{r_i\}$. Then, $\mathbf{B}_i$ is reconstructed as
\begin{eqnarray}
\mathbf{B}_i = \mathrm{Toeplitz}([1,r,\cdots,r^{d_i-1}]). \label{equ:B}
\end{eqnarray}
Our algorithm using this regularization (\ref{equ:r})-(\ref{equ:B}) is denoted by {\bf BSBL-FM(2)}.

\subsection{Remarks on $\beta$}

The parameter $\beta^{-1}$ is the noise variance in our model. It can be estimated by\cite{Zhang2012a}, 
\begin{equation}
	\beta = \frac{M}{\mathrm{Tr}[\bm{\Sigma}\bm{\Phi}^T\bm{\Phi}] + \norm{\ve{y}-\bm{\Phi}\bm{\mu}}_2^2}.
\end{equation}
However, the resulting updating rule is not robust due to the constructive and reconstruction nature\cite{Tipping2003,Babacan2012} of the proposed algorithm. In practice, people treat it as a regularizer and assign some specific values to it \footnote{For example, one can see this by examining the published codes of the algorithms in \cite{Ji2008,Babacan2012}.}. Similar to \cite{Ji2008}, we select $\beta^{-1}=10^{-6}$ in noiseless simulations, $\beta^{-1}=0.1\norm{y}_2^2$ in general noisy scenarios (e.g. $\text{SNR}<20$ dB), and $\beta^{-1}=0.01\norm{y}_2^2$ in high SNR scenarios (e.g. $\text{SNR}\geq20$ dB).

\subsection{The BSBL-FM algorithm}
The proposed algorithm (denoted as \textbf{BSBL-FM}) is given in Fig. \ref{algo:bsbl-fm}. 
\begin{figure}[h!]
\centering
\begin{algorithmic}[1]
    \algrule
\Procedure{BSBL-FM}{$\ve{y}$,$\bm{\Phi}$,$\eta$}
\State Outputs: $\ve{x},\bm{\Sigma}$
\State Initialize $\beta^{-1}= 0.01\norm{\ve{y}}_2^2$
\State Initialize $\{\ve{s}_i\}$, $\{\ve{q}_i\}$
\While{not converged}
\State Calculate $\ve{A}'_i= \ve{s}_i^{-1}(\ve{q}_i\ve{q}_i^T - \ve{s}_i)\ve{s}_i^{-1}, \forall i$
\State Calculate the block relevance $\gamma_i = \frac{1}{d_i} \mathrm{Tr}(\ve{A}'_i)$
\State Inferring Correlation Models $\ve{B}^*_i$ from  $\ve{A}'_i / \gamma_i$
\State Re-build $\ve{A}^*_i = \gamma_i\ve{B}^*_i$
\State Calculate $\Delta \mathcal{L}(i) = \mathcal{L}(\ve{A}^*_i) - \mathcal{L}(\ve{A}_i), \forall i$
\State Select the $\hat{i}$th block s.t. $\Delta\mathcal{L}(\hat{i})=\min\{\Delta\mathcal{L}(i)\}$
\State Re-calculate $\bm{\mu},\bm{\Sigma},\{\ve{s}_i\},\{\ve{q}_i\}$
\EndWhile
\EndProcedure
    \algrule
\end{algorithmic}
\caption{The Proposed BSBL-FM Algorithm.}
\label{algo:bsbl-fm}
\end{figure}
Within each iteration, it only updates the block signal that attributes to the deepest descent of $\mathcal{L}(i)$. The detailed procedures on re-calculation of $\bm{\mu},\bm{\Sigma},\{\ve{s}_i\},\{\ve{q}_i\}$ are similar to \cite{Tipping2003}. The algorithm terminates when the maximum change of the cost function is smaller than a threshold $\eta$. In the experiments thereafter, we set $\eta=1\mathrm{e}^{-4}$.

%
%
\section{Experiments}
In the experiments\footnote{Available on-line:
http://nudtpaper.googlecode.com/files/bsbl\_fm.zip} we compared the proposed algorithm with the state-of-the-art block based recovery algorithms. For comparison, two BSBL algorithms, i.e., BSBL-BO and BSBL-$\ell_1$ \cite{Zhang2012a}, were used (BSBL-$\ell_1$ used the Group Basis Pursuit \cite{van2008probing} in its inner loop). Besides, a variational inference based SBL algorithm (denoted by VBGS\cite{Babacan2012}) was selected. It used its default parameters. Model-CoSaMP\cite{Baraniuk2010} and Block-OMP\cite{Eldar2010} (given the true sparsity) were used as the benchmark in noiseless situations, while the Group Basis Pursuit \cite{van2008probing} was used as the benchmark in noisy situations.

The performance indexes were the normalized mean square error (NMSE) in noisy situations and the  success rate in noiseless situations. The NMSE was defined as 
$\norm{\ve{\hat{x}} - \ve{x}_{gen}}_2^2/\norm{\ve{x}_{gen}}_2^2$,
where $\ve{\hat{x}}$ was the estimate of the true signal $\ve{x}_{gen}$.
The success rate was defined as the
percentage of successful trials in total experiments (A successful trial
was defined the one when NMSE$\leq10^{-5}$).

In all the experiments except for the last one, the sensing matrix was a random Gaussian matrix, and it was generated in each trial of each experiment.
The computer used in the experiments had 2.5GHz CPU and 2G RAM.

\subsection{Empirical Phase Transition}
\label{subsec:phase}

In the first experiment, we studied the phase transitions of all the algorithms in exact
recovery of block sparse signals in noiseless situations. The phase transition curve \cite{donoho2009observed} is to show how the success rate is affected by the sparsity
level (defined as $\rho=K/M$, where $K$ is the total number of non-zero elements) and
indeterminacy (defined as $\delta=M/N$).

The generated signal consisted of $20$ blocks with the identical block size 25. The number of non-zero blocks varied from $1$ to $10$ while their locations were determined randomly. Each non-zero block was generated by
a multivariate Gaussian distribution $\mathcal{N}(\ve{0},\bm{\Sigma}_{\text{gen}})$ with $\mathbf{\Sigma}_{\text{gen}} \triangleq \mathrm{Toeplitz}([1,r,\cdots,r^{24}])$. The parameter $r$, which reflected the intra-block correlation level, was set to 0.95. The number of measurements
varied from $M=50$ to $M=250$.

\begin{figure}[!tbp]
\centering
\includegraphics[width=2.6in]{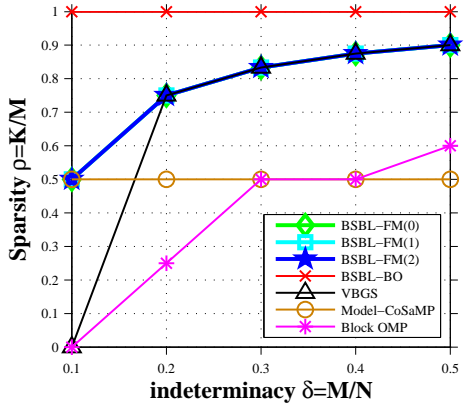}
\caption{Empirical 96\% phase transitions
of all algorithms.
Each point on the plotted phase transition curve
corresponds to the success rate larger than or equal to $0.96$.
Above the curve the success rate sharply drops.
}\label{fig:phase}
\end{figure}

The results averaged over 100 trials are shown in Fig. \ref{fig:phase}. 
Both BSBL-FM and BSBL-BO showed impressive phase transition performance. We see that as a greedy method, BSBL-FM performed better than VBGS, Model-CoSaMP and Block-OMP. 

\subsection{Performance in Noisy Environments with varying $N$}

This experiment was designed to show the advantage of our algorithm in speed. The signal consisted of $32$ blocks with identical block size, five of which were randomly located non-zero blocks. The length of the signal, $N$, was varied from $512$ to $2048$ with fixed indeterminacy ratio $M/N=0.5$.
The intra-block correlation level, i.e., $r$, of each block (generated as in Section \ref{subsec:phase}) was uniformly chosen from $0.8$ to $0.99$. 
The SNR, defined as $\text{SNR(dB)}\triangleq20\log_{10}(\norm{\bm{\Phi}\ve{x}_{gen}}_2/\norm{\mathbf{n}}_2)$, was fixed to $15$dB.
In this experiment we also calculated the oracle result, which was the least square estimate of $\mathbf{x}_{gen}$ given the true support.

\begin{figure}[!tbp]
\centering
\subfloat{\includegraphics[width=2.6in]{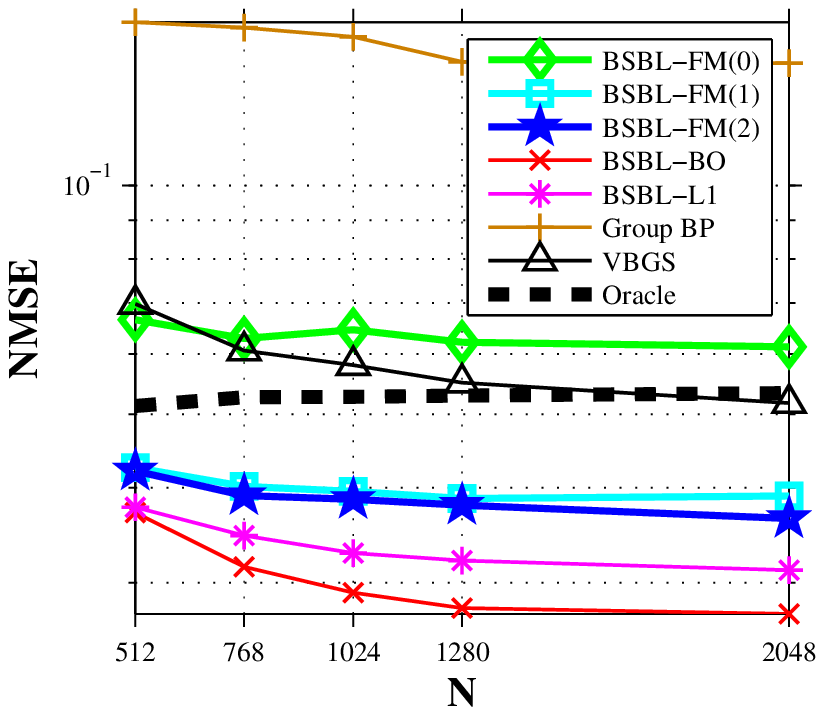}} \\
\subfloat{\includegraphics[width=2.6in]{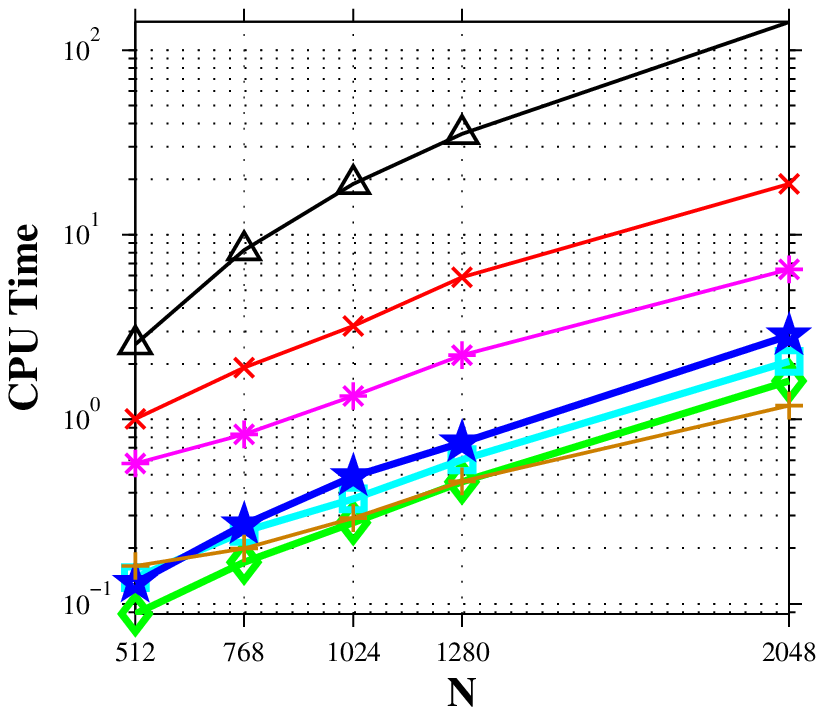}}
\caption{The comparison in NMSE and CPU Time with varying $N$.
}%
\label{fig:exp3}
\end{figure}

The results (Fig. \ref{fig:exp3}) show that the proposed algorithm, although the recovery performance was slightly poorer than BSBL-BO and BSBL-$\ell_1$, had the obvious advantage in speed. This implies that the proposed algorithm may be a better choice for large-scale problems. 
The BSBL-FM(1), BSBL-FM(2) and BSBL-BO even outperformed the oracle estimate, this may be due to that the oracle property utilized only the true support information while ignored the structure in signals (i.e., the intra-block correlation). 
Also, by comparing BSBL-FM(1) and BSBL-FM(2) to BSBL-FM(0), we can see its recovery performance was improved due to the exploitation of intra-block correlation.

\subsection{Application to Telemonitoring of Fetal Electrocardiogram}
Fetal electrocardiogram (FECG) telemonitoring via low energy wireless body-area networks \cite{Zhang_TBME2012b} is an important approach to monitor
fetus health state. BSBL, as an important branch of compressed sensing, has shown great promising in this application\cite{Zhang_TBME2012b}. 
Using BSBL, one can compress raw FECG recordings using a sparse binary matrix, i.e.,
\begin{equation}
\ve{y} = \bm{\Phi}\ve{x}
\label{equ:compressFECG}
\end{equation}
where $\mathbf{x}$ is a raw FECG recording, $\bm{\Phi}$ is the sparse binary matrix, and $\mathbf{y}$ is the compressed data. It have been showed\cite{mamaghanian2011compressed} that using a sparse binary matrix as the sensing matrix can greatly reduce the energy consumption while achieving competitive compression ratio. Then $\mathbf{y}$ is sent to a remote computer. In this computer BSBL algorithms can recover the raw FECG recordings with high accuracy such that Independent Component Analysis (ICA) decomposition \cite{hyvarinen1999fast} on the recovered recordings keeps high fidelity (and a clean FECG is presented after the ICA decomposition).

Here we repeated the experiment in Section III.B in \cite{Zhang_TBME2012b} \footnote{Available on-line: https://sites.google.com/site/researchbyzhang/bsbl}. 
using the same dataset, the same sensing matrix (a sparse binary matrix with the size $256\times 512$ and each column consisting of 12 entries of $1$s with random locations), and the same block partition ($d_i=32(\forall i)$).

We compared our algorithm BSBL-FM with VBGS, Group-BP and BSBL-BO. 
All the algorithms first recovered the discrete cosine transform (DCT) coefficients $\boldsymbol{\theta}$ of the recordings according to
\begin{equation}
\mathbf{y}=(\mathbf{\Phi D}) \boldsymbol{\theta}
\end{equation}
using $\mathbf{y}$ and $\mathbf{\Phi D}$, where $\mathbf{D}$ was the basis of the DCT transform such that $\mathbf{x} = \mathbf{D}\boldsymbol{\theta}$. Then we reconstructed the original raw FECG recordings according to $\mathbf{x} = \mathbf{D}\boldsymbol{\theta}$ using $\mathbf{D}$ and $\boldsymbol{\theta}$.

\begin{figure}[!tbp]
\centering
\includegraphics[width=2.6in]{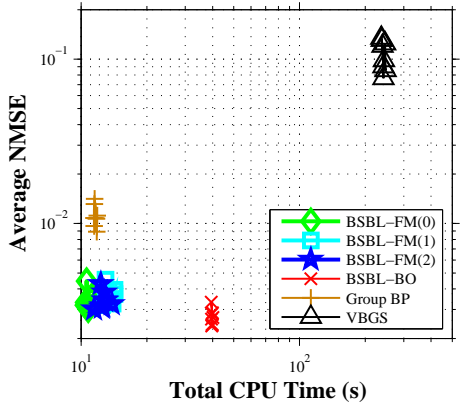}
\caption{Average NMSE and the total CPU time in recovery of the FECG recordings. Each data point on a curve corresponds to the average NMSE and total CPU time to reconstruct the data of a channel in the raw FECG recordings.}
\label{tab:fecg}
\end{figure}

\begin{figure}[!tbp]
\centering
\subfloat{\includegraphics[width=1.7in]{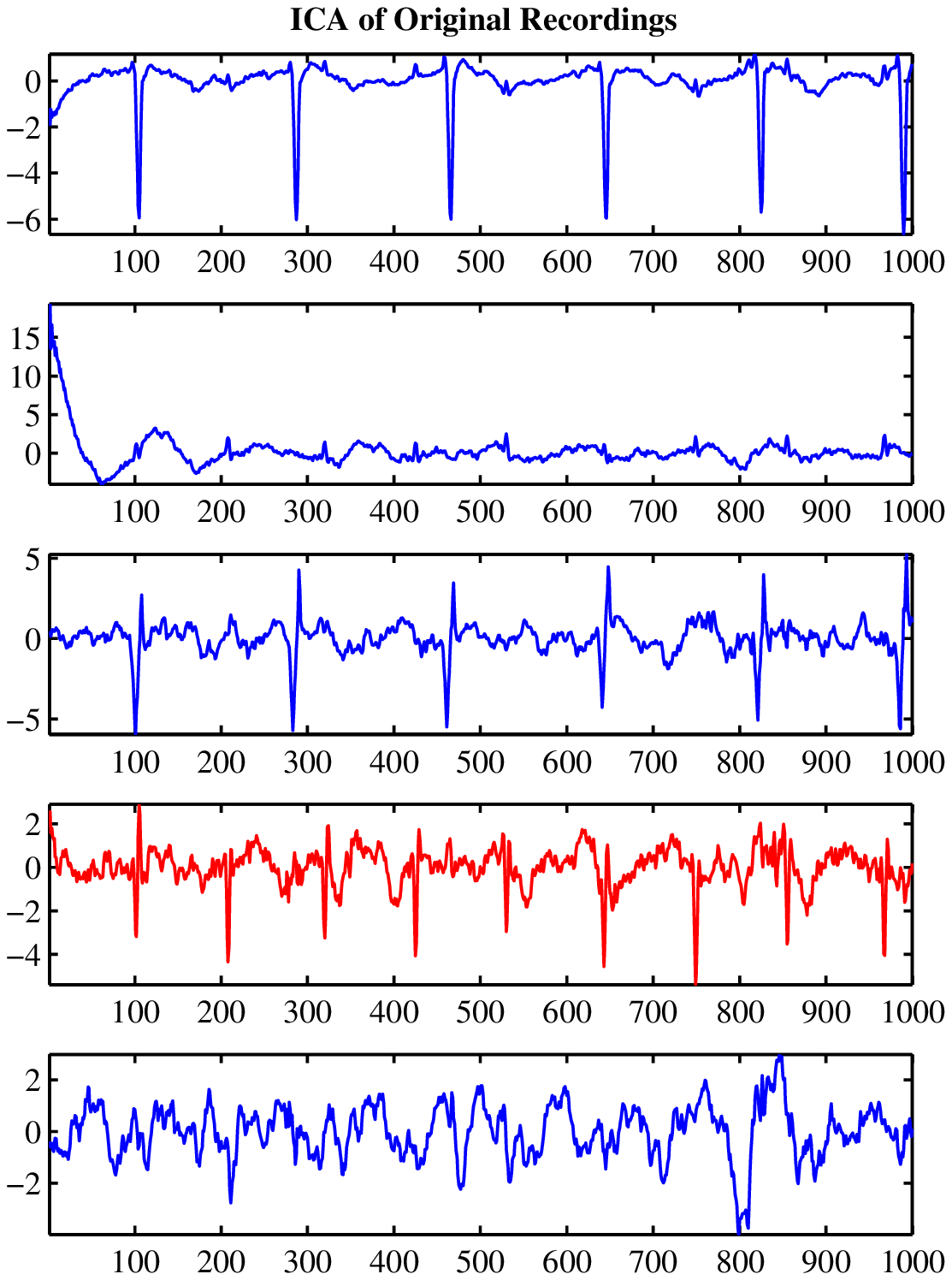}}
\subfloat{\includegraphics[width=1.7in]{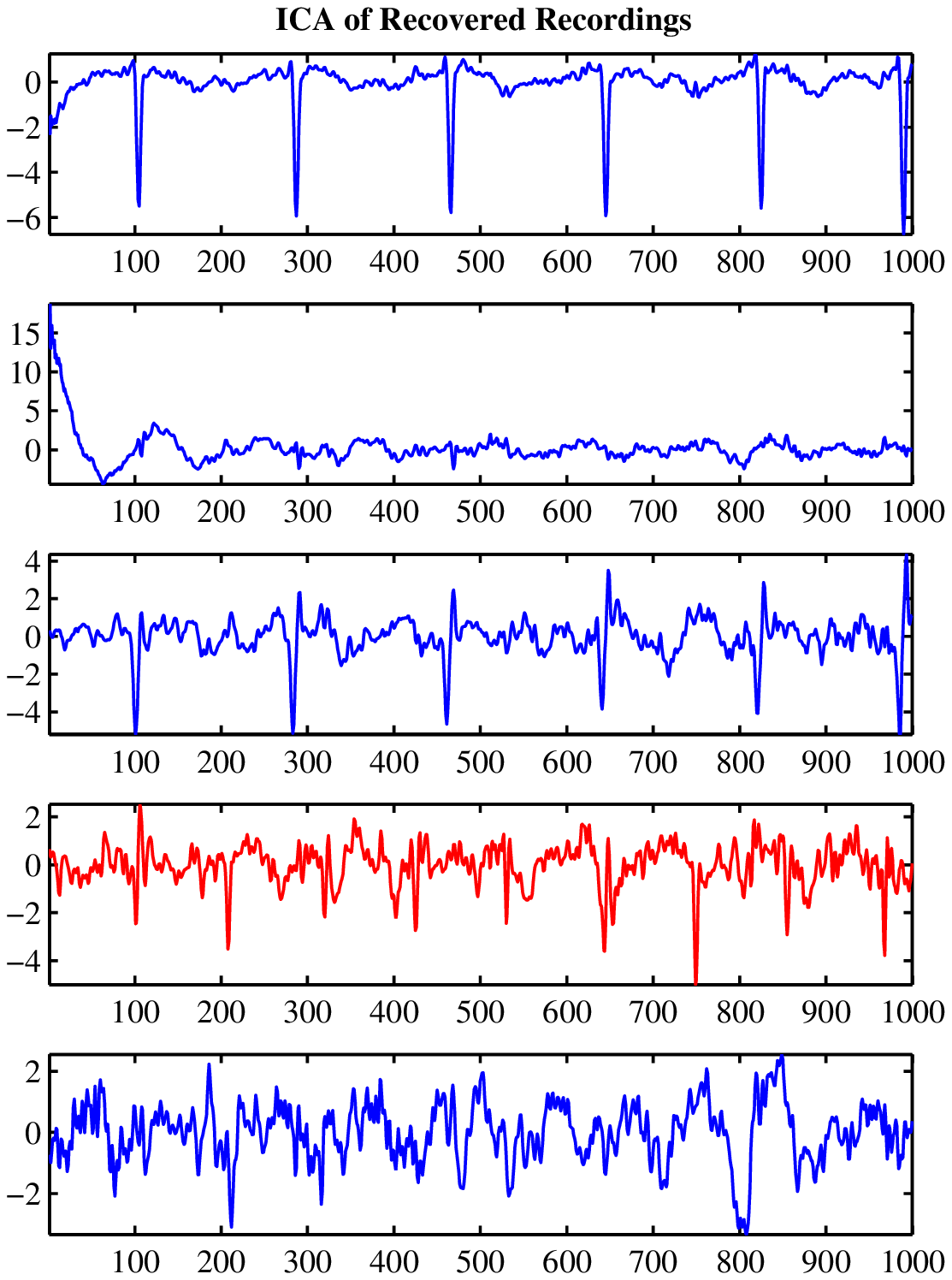}}
\caption{ICA decomposition of the original dataset and the recovered dataset by BSBL-FM(0) (only the first $1000$ sampling points of the datasets are shown). The fourth ICs are the extracted FECGs.
}
\label{fig:fecg}
\end{figure}

The NMSE measured on the recovered FECG recordings is shown in Fig. \ref{tab:fecg}. We can see although BSBL-FM had slightly poorer recovery accuracy than BSBL-BO, it had much faster speed. In fact, the ICA decomposition on the recovered recordings by BSBL-FM also presented a clean FECG (see Fig. \ref{fig:fecg}), and the decomposition was almost the same as the ICA decomposition on the original recordings. In this experiment we noticed that VBGS took long time to recover the FECG recordings, and had the largest NMSE. Besides, the ICA decomposition on its recovered recordings didn't present the clean FECG. This reason may be due to the fact that the DCT coefficients of the raw FECG recordings are not sufficiently sparse,  and recovering these less-sparse coefficients is very difficult for non-BSBL algorithms\cite{Zhang_TBME2012b}. This experiment shows the robustness and the speed efficiency of the proposed algorithm applied to real-life applications.

%

%
%
\section{Conclusion}

In this paper, we proposed a fast BSBL algorithm that can exploit both the block sparsity and the intra-block correlation of the signal. Experiments showed that it significantly outperforms non-BSBL algorithms, and has close recovery performance as existing BSBL algorithms, but is the fastest among the BSBL algorithms.



%



\ifCLASSOPTIONcaptionsoff
  \newpage
\fi



\bibliographystyle{IEEEtran}
\bibliography{bsbl}

\begin{thebibliography}{10}
\providecommand{\url}[1]{#1}
\csname url@samestyle\endcsname
\providecommand{\newblock}{\relax}
\providecommand{\bibinfo}[2]{#2}
\providecommand{\BIBentrySTDinterwordspacing}{\spaceskip=0pt\relax}
\providecommand{\BIBentryALTinterwordstretchfactor}{4}
\providecommand{\BIBentryALTinterwordspacing}{\spaceskip=\fontdimen2\font plus
\BIBentryALTinterwordstretchfactor\fontdimen3\font minus
  \fontdimen4\font\relax}
\providecommand{\BIBforeignlanguage}[2]{{%
\expandafter\ifx\csname l@#1\endcsname\relax
\typeout{** WARNING: IEEEtran.bst: No hyphenation pattern has been}%
\typeout{** loaded for the language `#1'. Using the pattern for}%
\typeout{** the default language instead.}%
\else
\language=\csname l@#1\endcsname
\fi
#2}}
\providecommand{\BIBdecl}{\relax}
\BIBdecl

\bibitem{Candes2008a}
E.~Candes and M.~Wakin, ``An introduction to compressive sampling,''
  \emph{Signal Processing Magazine, IEEE}, vol.~25, no.~2, pp. 21 --30, march
  2008.

\bibitem{Baraniuk2010}
R.~G. Baraniuk, V.~Cevher, M.~F. Duarte, and C.~Hegde, ``Model-based
  compressive sensing,'' \emph{IEEE Transactions on Signal Processing}, vol. 56
  (4), pp. 1982--2001, 2010.

\bibitem{Zhang2012a}
Z.~Zhang and B.~Rao, ``Extension of {SBL} algorithms for the recovery of block
  sparse signals with intra-block correlation,'' \emph{Signal Processing, IEEE
  Transactions on}, vol.~61, no.~8, pp. 2009--2015, 2013.

\bibitem{Yuan2006}
M.~Yuan and Y.~Lin, ``Model selection and estimation in regression with grouped
  variables,'' \emph{J. R. Statist. Soc. B}, vol.~68, pp. 49--67, 2006.

\bibitem{Babacan2012}
S.~D. Babacan, S.~Nakajima, and M.~N. Do, ``Bayesian group-sparse modeling and
  variational inference,'' \emph{Submitted to IEEE Transactions on Signal
  Processing}, 2012.

\bibitem{Zhang2011}
Z.~Zhang and B.~D. Rao, ``Sparse signal recovery with temporally correlated
  source vectors using sparse bayesian learning,'' \emph{IEEE Journal of
  Selected Topics in Signal Processing}, vol.~5, no.~5, pp. 912--926, 2011.

\bibitem{Zhang_TBME2012b}
Z.~Zhang, T.-P. Jung, S.~Makeig, and B.~Rao, ``Compressed sensing for
  energy-efficient wireless telemonitoring of noninvasive fetal {ECG} via block
  sparse bayesian learning,'' \emph{Biomedical Engineering, IEEE Transactions
  on}, vol.~60, no.~2, pp. 300--309, 2013.

\bibitem{Tipping2003}
M.~E. Tipping and A.~C. Faul, ``Fast marginal likelihood maximisation for
  sparse bayesian models,'' in \emph{Proceedings of the Ninth International
  Workshop on Artificial Intelligence and Statistics}, C.~M. Bishop and B.~J.
  Frey, Eds., Key West, FL, 2003, pp. 3--6.

\bibitem{Eldar2010}
Y.~C. Eldar, P.~Kuppinger, and H.~Bolcskei, ``Block-sparse signals: uncertainty
  relations and efficient recovery,'' \emph{IEEE Transaction on Signal
  Processing}, vol. 58(6), pp. 3042--3054, 2010.

\bibitem{Tipping2001}
M.~E. Tipping, ``Sparse bayesian learning and the relevance vector machine,''
  \emph{Journal of Machine Learning Research}, vol.~1, pp. 211--244, 2001.

\bibitem{Ji2008}
S.~Ji, Y.~Xue, and L.~Carin, ``Bayesian compressive sensing,'' \emph{IEEE
  Transactions on Signal Processing}, vol. 56 (6), pp. 2346--2356, 2008.

\bibitem{van2008probing}
E.~Van Den~Berg and M.~Friedlander, ``Probing the pareto frontier for basis
  pursuit solutions,'' \emph{SIAM Journal on Scientific Computing}, vol.~31,
  no.~2, pp. 890--912, 2008.

\bibitem{donoho2009observed}
D.~Donoho and J.~Tanner, ``Observed universality of phase transitions in
  high-dimensional geometry, with implications for modern data analysis and
  signal processing,'' \emph{Philosophical Transactions of the Royal Society
  A}, vol. 367, no. 1906, pp. 4273--4293, 2009.

\bibitem{mamaghanian2011compressed}
H.~Mamaghanian, N.~Khaled, D.~Atienza, and P.~Vandergheynst, ``Compressed
  sensing for real-time energy-efficient {ECG} compression on wireless body
  sensor nodes,'' \emph{Biomedical Engineering, IEEE Transactions on}, vol.~58,
  no.~9, pp. 2456--2466, 2011.

\bibitem{hyvarinen1999fast}
A.~Hyvarinen, ``Fast and robust fixed-point algorithms for independent
  component analysis,'' \emph{Neural Networks, IEEE Transactions on}, vol.~10,
  no.~3, pp. 626--634, 1999.

\end{thebibliography}
%



%






\end{document}